\documentclass[preprint,preprintnumbers, prd, floatfix, superscriptaddress,nofootinbib] {revtex4-1}
\usepackage{epsfig}
\usepackage{subfigure}
\usepackage{dcolumn}
\usepackage{bm}
\usepackage[usenames ,dvipsnames]{xcolor}
\usepackage{slashed}
\usepackage{graphicx,color}

\begin{document}
\title{Angular asymmetries in $B\to\Lambda\bar p M$ decays}
\author{Xiaotao Huang}
\affiliation{The Institute for Advanced Studies, Wuhan University, Wuhan 430072, China}

\author{Yu-Kuo Hsiao}
\email{Corresponding author: yukuohsiao@gmail.com}
\affiliation{School of Physics and Information Engineering, Shanxi Normal University, 
Taiyuan, 030031, China}

\author{Jike Wang}
\email{Corresponding author: Jike.Wang@whu.edu.cn}
\affiliation{The Institute for Advanced Studies, Wuhan University, Wuhan 430072, China}

\author{Liang Sun}
\email{Corresponding author: sunl@whu.edu.cn}
\affiliation{School of Physics and Technology, Wuhan University, Wuhan 430072, China}
\date{\today}

\begin{abstract}
The forward-backward angular asymmetry (${\cal A}_{FB}$) for $\bar B^0\to \Lambda\bar p\pi^+$
measured by Belle has presented an experimental value in the range of $-30\%$ to $-50\%$. In our study,
we find that ${\cal A}_{FB}[\bar B^0\to\Lambda\bar p \pi^+(B^-\to \Lambda\bar p\pi^0)]$
can be as large as $(-14.6^{+0.9}_{-1.5}\pm 6.9)\%$. In addition, we present
${\cal A}_{FB}[\bar B^0\to\Lambda\bar p \rho^+(B^-\to \Lambda\bar p\rho^0)]
=(4.1^{+2.8}_{-0.7}\pm 2.0)\%$ as the first prediction involving a vector meson
in the charmless $B\to{\bf B\bar B'}M$ decays.
While ${\cal A}_{FB}(B\to\Lambda\bar p M)$ indicates an angular correlation 
caused by the rarely studied baryonic form factors in the timelike region,
LHCb and Belle~II are capable of performing experimental examinations.
\end{abstract}

\maketitle
\section{introduction}
For the three-body baryonic $B\to{\bf B\bar B'}M_{(c)}$ decays
with $M_{(c)}$ denoting a (charmed) meson, 
the threshold effect has been commonly observed with a rapidly raising peak 
around the threshold area of $m_{\bf B\bar B'}\simeq m_{\bf B}+m_{\bf \bar B'}$
in the dibaryon invariant mass ($m_{\bf B\bar B'}$) spectrum~\cite{Belle:2005mke}.
It is regarded to enhance the branching fraction of 
$B\to{\bf B\bar B'}M_{(c)}$ $[{\cal B}(B\to{\bf B\bar B'}M_{(c)})]$~\cite{Hou:2000bz,Suzuki:2006nn}, 
such as ${\cal B}(B^-\to p\bar p M)\sim 10^{-6}$ 
with $M=(\pi^-,K^-,K^{*-})$~\cite{BaBar:2005sdl,Belle:2007oni,LHCb:2014nix,Belle:2008zkc} and 
${\cal B}(\bar B^0\to p\bar p D^{(*)0})\sim 10^{-4}$~\cite{Belle:2002fay,BaBar:2011zka}. On the other hand,
${\cal B}(\bar B^0\to p\bar p)$ is as small as $10^{-8}$~\cite{LHCb:2017swz,pdg}, 
whose suppression reflects the fact that in the two-body baryonic $B$ decays
the $\bf B\bar B'$ formation with $m_{\bf B\bar B'}\sim m_B$ is away from the threshold area. 
Theoretically, the baryonic form factors that parameterize the dibaryon formation
have been used to describe the threshold effect~\cite{Brodsky:1973kr,Lepage:1979za,
Lepage:1980fj,Brodsky:1973kr,Brodsky:2003gs,Chua:2002wn,Geng:2006wz,Chua:2001vh},
such that ${\cal B}(B\to{\bf B\bar B'}M_{(c)})$ can be explained.

The partial branching fraction can be a function of $\cos\theta_{\bf B(\bar B')}$, 
where $\theta_{\bf B(\bar B')}$ is the angle between
the (anti-)baryon and meson moving directions in the $\bf B\bar B'$ rest frame.
It leads to the forward-backward angular asymmetry:
${\cal A}_{FB}(B\to{\bf B\bar B'}M_{(c)}) 
\equiv ({\cal B}_+ -{\cal B}_-)/({\cal B}_+ +{\cal B}_-)$, where
${\cal B}_{+}={\cal B}(\cos\theta_{\bf B(\bar B')}>0)$ and 
${\cal B}_{-}={\cal B}(\cos\theta_{\bf B(\bar B')}<0)$.
The forward-backward asymmetries have been found 
in several decay channels~\cite{Belle:2007lbz,LHCb:2014nix,
Belle:2007oni,Chang:2015fja}, of which the interpretations have caused 
theoretical difficulties~\cite{Geng:2006wz,Tsai,Hsiao:2016amt}.
This indicates that the dibaryon production in $B\to{\bf B\bar B'}M$ 
has not been fully understood~\cite{Huang:2021qld}.

One has measured the angular asymmetries of 
$B\to\Lambda\bar p M_{(c)}$ versus $\cos\theta_{\bar p}$ 
in Refs.~\cite{Chang:2015fja,Belle:2007lbz}, that is,
\begin{eqnarray}\label{ang_data}
&&
{\cal A}_{FB}(\bar B^0\to \Lambda\bar p D^+)
=(-8\pm 10)\%\,,\nonumber\\
&&
{\cal A}_{FB}(\bar B^0\to \Lambda\bar p D^{*+})
=(55\pm 17)\%\,,\nonumber\\
&&
{\cal A}_{FB}(\bar B^0\to \Lambda \bar p\pi^+)=(-41\pm 11\pm 3)\%\,\,,\nonumber\\
&&
{\cal A}_{FB}(B^-\to \Lambda \bar p\pi^0)=(-16\pm 18\pm 3)\%\,.
\end{eqnarray}
According to the calculation~\cite{Hsiao:2016amt},
${\cal A}_{FB}(\bar B^0\to \Lambda\bar p D^{+})=(-3.0\pm 0.2)\%$ is in agreement with the data;
${\cal A}_{FB}(\bar B^0\to \Lambda\bar p D^{*+})=(15.0\pm 0.0)\%$
presents a sizeable asymmetry despite of two standard deviation compared to the data in Eq.~(\ref{ang_data}). 
For $B\to\Lambda\bar p\pi$, different angular observables have been studied. 
One is the angular distribution of the cascade $B\to \bar p\pi(\Lambda\to)p\pi^-$ decay
versus $\cos\theta$~\cite{Chua:2002yd}, where $\theta$ denotes the angle between
proton and $B$ meson moving directions in the $\Lambda$ rest frame.
Subsequently, C.K.~Chua and W.S.~Hou of Ref.~\cite{Chua:2002yd} demonstrate that 
$\Lambda$ is dominantly a left-handed state~\cite{Suzuki:2002cp}, 
consistent with the experimental result in Ref.~\cite{BaBar:2009ess}.
The other study is from Ref.~\cite{Tsai}, 
where S.Y.~Tsai presents ${\cal A}_{FB}(\bar B^0\to \Lambda \bar p\pi^+)\simeq 0$,
not verified by the later measurement as in Eq.~(\ref{ang_data}).
Clearly, it indicates a sizeable angular correlation to be discovered
in the charmless baryonic $B\to{\bf B\bar B'}M$ decays. 
Moreover, the isospin symmetry should lead to
${\cal B}(\bar B^0\to \Lambda \bar p\pi^+)=2{\cal B}(B^-\to \Lambda \bar p\pi^0)$ and
${\cal A}_{FB}(\bar B^0\to \Lambda \bar p\pi^+)={\cal A}_{FB}(B^-\to \Lambda \bar p\pi^0)$~\cite{Chua:2002yd},
which seem to disagree with 
${\cal B}(\bar B^0\to \Lambda \bar p\pi^+)=(3.14\pm 0.29)\times 10^{-6}$~\cite{pdg,BaBar:2009ess,Belle:2007lbz},
${\cal B}(B^-\to \Lambda \bar p\pi^0)=(3.0^{+0.7}_{-0.6})\times 10^{-6}$~\cite{pdg,Belle:2007lbz},
and angular asymmetries in Eq.~(\ref{ang_data})~\cite{Belle:2007lbz}.
This suggests a possible isospin symmetry violation to be tested.

Compared to the charmful $\bar B^0\to \Lambda\bar p D^{(*)+}$ decay channels,
where the $\Lambda\bar p$ formation is from the (axial)vector current,
the penguin-dominant $B\to\Lambda\bar p \pi$ decay 
has an additional contribution from the (pseudo)scalar current. Consequently, 
there might exist an interference between the (axial)vector and (pseudo)scalar currents,
which can cause a possible angular asymmetry.
Therefore, we propose to investigate $B\to\Lambda\bar p\pi$, along with 
the rarely studied baryonic form factors in the timelike region.
We will also study ${\cal A}_{FB}(B\to\Lambda\bar p \rho)$, 
which can be the first prediction involving a vector meson 
in the charmless $B\to{\bf B\bar B'}M$ decays.
The isospin relations will be discussed.
%
\begin{figure}[t!]
\centering
\includegraphics[width=2.8in]{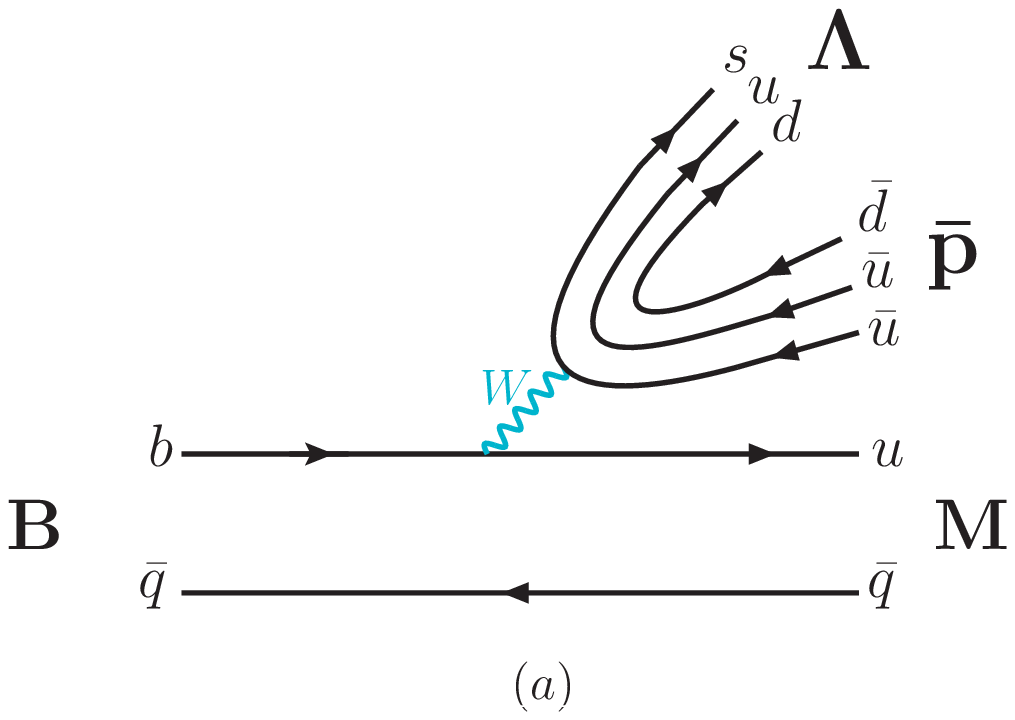}
\includegraphics[width=2.8in]{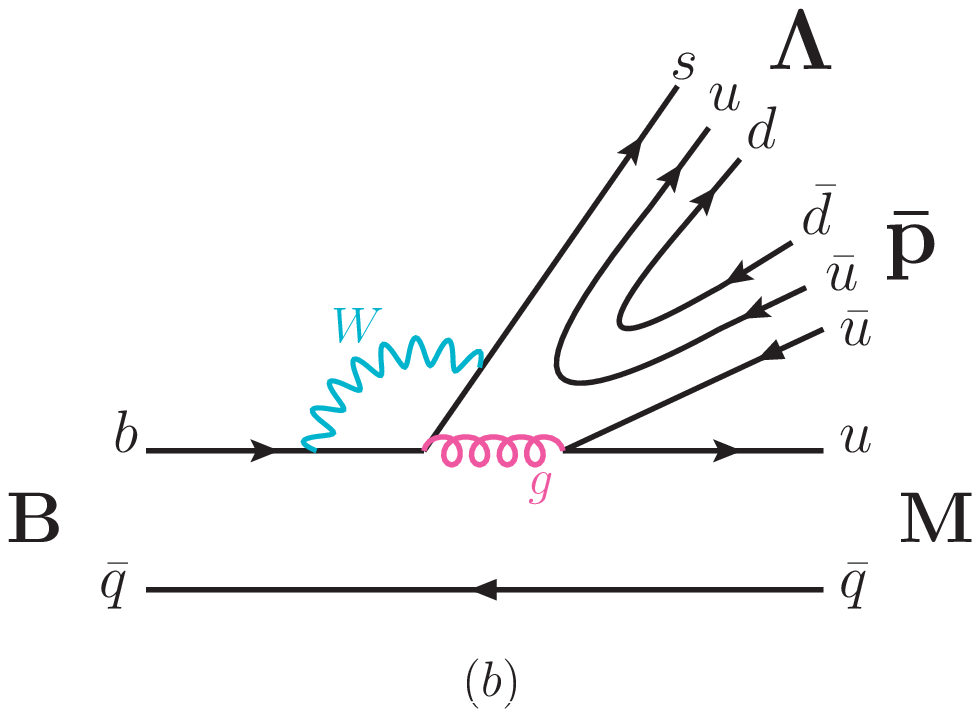}
\caption{Feynman diagrams for $B\to \Lambda\bar p M$ 
that depict (a) tree and (b) penguin-level processes, 
respectively.}\label{fig1}
\end{figure}
%
\section{Formalism}
According to Fig.~\ref{fig1}, where the decay process is drawn with
the $B$ meson transition to a meson, along with the dibaryon production,
the amplitude of $B\to\Lambda \bar p M$ 
can be factorized as~\cite{Chua:2002yd,Tsai,Geng:2005wt,Geng:2006yk,Geng:2011pw}
\begin{eqnarray}\label{amp1}
&&{\cal M}(B\to\Lambda \bar p M)=\nonumber\\
&&\frac{G_F}{\sqrt 2}
\bigg[\alpha_1\langle \Lambda \bar p|(\bar s u)_{V-A}|0\rangle\langle M|(\bar u b)_{V-A}|B(b\bar q)\rangle
+\alpha_6\langle \Lambda \bar p|(\bar s u)_{S+P}|0\rangle\langle M|(\bar u b)_{S-P}|B(b\bar q)\rangle\bigg]\;,
\end{eqnarray}
with $q=d$ and $u$ for $\bar B^0\to\Lambda\bar p\pi^+(\rho^+)$ and
$B^-\to\Lambda\bar p\pi^0(\rho^0)$, respectively,
where $G_F$ is the Fermi constant, 
$(\bar q_iq_j)_{V-A}=\bar q_i\gamma_\mu(1-\gamma_5)q_j$, 
$(\bar q_iq_j)_{S\pm P }=\bar q_i(1\pm\gamma_5)q_j$, and
$|0\rangle$ in $\langle \Lambda \bar p|(\bar s u)|0\rangle$
denotes the vacuum state. We define
$\alpha_1=V_{ub}V_{us}^* a_1-V_{tb}V_{ts}^*a_4$ and
$\alpha_6=V_{tb}V_{ts}^*2a_6$, where
$V_{q_iq_j}$ are the Cabibbo–Kobayashi–Maskawa (CKM) matrix elements,
and the parameters $a_{1,4,6}$ consist of the effective Wilson coefficients $c_i^{eff}$~\cite{ali} , 
given by
\begin{eqnarray}\label{a146}
a_1=c_1^{eff}+\frac{1}{N_c}c_2^{eff}\,,\;
a_4=c_4^{eff}+\frac{1}{N_c}c_3^{eff}\,,\;
a_6=c_6^{eff}+\frac{1}{N_c}c_5^{eff}\,,
\end{eqnarray}
with $N_c$ the color number. 

In Eq.~(\ref{amp1}), the matrix elements of the $B$ to $\pi(\rho)$ transition 
can be written as~\cite{MS} 
\begin{eqnarray}\label{dc}
&&
\langle \pi|(\bar u b)_{V-A}|B\rangle=
p^\mu F_{\pi1}+\frac{m_B^2-m_\pi^2}{t}q^\mu (F_{\pi0}-F_{\pi1})\,,\nonumber\\
&&
\langle \rho|(\bar u b)_{V-A}|B\rangle=
\epsilon_{\mu\nu\alpha\beta}
\varepsilon^{\ast\nu}p_B^{\alpha}p_\rho^{\beta}\frac{2V_1}{m_+}\,\nonumber\\
&&
-i\bigg\{\bigg[\varepsilon^\ast_\mu-\frac{\varepsilon^\ast\cdot q}{t}q_\mu\bigg](m_+)A_1
+\frac{\varepsilon^\ast\cdot q}{t}q_\mu(2m_\rho)A_0
-\bigg[p_\mu-\frac{m_B^2-m_\rho^2}{t}q_\mu \bigg]
(\varepsilon^\ast\cdot q)\frac{A_2}{m_+}\bigg\}\,,
\end{eqnarray}
with $p^\mu=(p_B+p_M)^\mu$, $q^\mu=(p_B-p_M)^\mu$,
$m_+=m_B+m_M$, $t\equiv q^2$, and $\varepsilon^\ast_\mu$ 
defined as the polarization four-vector of the $\rho$ meson,  
where $F_A=(F_{\pi1},V_1,A_0)$ and $F_B=(F_{\pi0},A_{1,2})$
are the mesonic form factors. Using the equation of motion, we obtain 
$\langle \pi|(\bar u b)_{S-P}|B\rangle=(p\cdot q/m_b)F_0^{\pi}$ and
$\langle \rho|(\bar u b)_{S-P}|B\rangle=2i(m_\rho/m_b)A_0\varepsilon^\ast\cdot q$.
The form factor $F_{A(B)}$ can be given in the three-parameter representation~\cite{MS}:
\begin{eqnarray}\label{form2}
&&F_A(t)=
\frac{F_A(0)}{(1-\frac{t}{M_A^2})
(1-\frac{\sigma_{1} t}{M_A^2}+\frac{\sigma_{2} t^2}{M_A^4})}\,,\;\nonumber\\
&&F_B(t)=
\frac{F_B(0)}{1-\frac{\sigma_{1} t}{M_B^2}+\frac{\sigma_{2} t^2}{M_B^4}}\,,
\end{eqnarray}
where $F_{A(B)}(0)$ is the form factor at the zero momentum transfer squared ($t=0$), 
$\sigma_{1,2}$ the parameters, and $M_{A(B)}$ the pole mass.
One has determined $F_{A(B)}(0)$, $\sigma_{1,2}$ and $M_{A(B)}$ with the results of the model calculation, 
such that the momentum transfer squared dependences for $F_{A(B)}$ can be described.

For the baryon-pair production, 
the matrix elements are in the forms of~\cite{Chua:2002wn,Geng:2005wt,
Chua:2002yd,Hsiao:2014zza,Hsiao:2019wyd}
\begin{eqnarray}\label{timelikeF}
\langle {\bf B}{\bf\bar B'}|\bar q\gamma_\mu q'|0\rangle
&=&
\bar u\bigg[F_1\gamma_\mu+\frac{F_2}{m_{\bf B}+m_{\bf \bar B'}}i\sigma_{\mu\nu}q_\mu\bigg]v\;,\nonumber\\
\langle {\bf B}{\bf\bar B'}|\bar q\gamma_\mu \gamma_5 q'|0\rangle
&=&\bar u\bigg[g_A\gamma_\mu+\frac{h_A}{m_{\bf B}+m_{\bf \bar B'}}q_\mu\bigg]\gamma_5 v\,,\nonumber\\
\langle {\bf B}{\bf\bar B'}|\bar q q'|0\rangle &=&f_S\bar uv\;,\nonumber\\
 \langle {\bf B}{\bf\bar B'}|\bar q\gamma_5 q'|0\rangle &=&g_P\bar u \gamma_5 v\,,
\end{eqnarray}
where the spinor $u(v)$ represents the spin-1/2 (anti-)baryon state,
$F_{\bf B\bar B'}=(F_{1,2},g_A,h_A)$ and $(f_S,g_P)$
are the baryonic form factors in the timelike region.
In the approach of pQCD counting rules, one derives that 
$F_{\bf B\bar B'}\propto (\alpha_s/t)^n$~\cite{Brodsky:1973kr,Lepage:1979za,
Lepage:1980fj,Brodsky:1973kr,Brodsky:2003gs}, 
where $n$ is to account for the gluon propagators 
that attach the valence quarks in $\bf B\bar B'$. Besides,
$\alpha_s(t)=(4\pi/\beta_0)[\text{ln}(t/\Lambda_0^2)]^{-1}$
is the running coupling constant in the strong interaction~\cite{Lepage:1980fj},
with the parameter $\beta_0\equiv 11-2n_f/3$,  the flavor number $n_f=3$, and the scale factor $\Lambda_0=0.3$~GeV. 
Subsequently, $(F_1,g_A,f_S,g_P)$ correspond to $n=2$; however,
$F_2$ and $h_A$ need an additional gluon to flip the chirality, such that $n=3$.
Explicitly, we present $F_{\bf B\bar B'}$ as~\cite{Brodsky:1973kr,Brodsky:2003gs,
Chua:2002wn,Geng:2006wz}
\begin{eqnarray}\label{timelikeF2}
&&(F_1,g_A)=
\frac{(C_{F_1},C_{g_A})}{t^2}\ln\bigg(\frac{t}{\Lambda_0^2}\bigg)^{-\gamma}\,,\nonumber\\
&&(f_S,g_P)=
\frac{(C_{f_S},C_{g_P})}{t^2}\ln\bigg(\frac{t}{\Lambda_0^2}\bigg)^{-\gamma}\,,\nonumber\\
&&
(F_2,h_A)=\frac{(C_{F_2},C_{h_A})}{t^3}\ln\bigg(\frac{t}{\Lambda_0^2}\bigg)^{-\gamma'}\,,
\end{eqnarray}
with $\gamma^{(\prime)}=2.148(3.148)$.\\

Using the $SU(2)$ helicity $[SU(2)_h]$ symmetry, 
$F_1$ and $g_A$ can be related. To this end,
we parameterize $\langle {\bf B}_{R+L}|J^{R}_\mu|{\bf B}'_{R+L}\rangle$ 
in the spacelike region as~\cite{Lepage:1979za,Hsiao:2014zza}
\begin{eqnarray}\label{Gff1}
\langle {\bf B}_{R+L}|J_\mu^R|{\bf B}'_{R+L}\rangle=
\bar u\bigg[\gamma_\mu \frac{1+\gamma_5}{2}F_R+\gamma_\mu \frac{1-\gamma_5}{2}F_L\bigg]u\,,
\end{eqnarray}
where $J^R_\mu=(V_\mu+A_\mu)/2$ is a right-handed chiral current,
$|{\bf B}^{(\prime)}_{R+L}\rangle\equiv |{\bf B}^{(\prime)}_R\rangle+|{\bf B}^{(\prime)}_L\rangle$,
and $F_{R,L}$ the chiral form factors. Furthermore,
we define $Q\equiv J^R_0$ as the chiral charge 
to act on the valence quark $q_i$ in ${\bf B}'(q_1 q_2 q_3)$, 
such that one transforms ${\bf B'}$ into ${\bf B}$.
With the chirality that is regarded as the helicity at $t\to\infty$,
the helicity of $q_i$ can be (anti-)parallel $[||(\overline{||})]$ to the helicity of ${\bf B}'$,
such that we denote the chiral charge for $q_i$ 
by $Q_{||(\overline{||})}(i)$ ($i=1,2,3$).
Thus, we obtain~\cite{Lepage:1979za,Hsiao:2014zza}
\begin{eqnarray}\label{Gff2}
(F_R,F_L)&=&
(e^R_{||}F_{||}+e^R_{\overline{||}}F_{\overline{||}},e^L_{||}F_{||}+e^L_{\overline{||}}F_{\overline{||}})\,,
\nonumber\\
e^R_{||(\overline{||})}
&=&\Sigma_i \langle {\bf B}_R|Q_{||(\overline{||})}(i)|{\bf B}'_R\rangle\,,\nonumber\\
e^L_{||(\overline{||})}
&=&\Sigma_i \langle {\bf B}_L|Q_{||(\overline{||})}(i)|{\bf B}'_L\rangle\,,
\end{eqnarray}
where $F_{||(\overline{||})}\equiv$ $C_{||(\overline{||})}/t^2[\ln(t/\Lambda^2)]^{-\gamma}$.
This results in
$C_{F_1}(C_{g_A})=(e^R_{||}\pm e^L_{||})C_{||}+(e^R_{\overline{||}}\pm e^L_{\overline{||}})C_{\overline{||}}$.
In the crossing symmetry, since
the spacelike form factors can be seen to behave as the timelike ones,
the derivation can be applied to
those in Eqs.~(\ref{timelikeF}, \ref{timelikeF2}).
Similarly, we relate $f_S$ and $g_P$.
It is hence obtained that 
\begin{eqnarray}\label{timelikeF3}
&&
(C_{F_1},C_{g_A})=\sqrt\frac{3}{2}(C_{||},C_{||}^*)\,,\;
(C_{f_S},C_{g_P})=-\sqrt\frac{3}{2}(\bar C_{||},\bar C_{||}^*)\,,\;\;
\text{for $\langle \Lambda\bar p|(\bar s u)_{V,A,S,P}|0\rangle$}\,,\nonumber\\
&&
(C_{F_1},C_{g_A})=\frac{1}{3}(4C_{||}-C_{\overline{||}},4C^*_{||}+C^*_{\overline{||}})\,,\;\;
\text{for $\langle n\bar p|(\bar d u)_{V,A}|0\rangle$}\,,
\end{eqnarray}
with $C_{||(\overline{||})}^*\equiv C_{||(\overline{||})}+\delta C_{||(\overline{||})}$ and 
$\bar C_{||}^*\equiv \bar C_{||}+\delta \bar C_{||}$, where the second line 
for $\langle n\bar p|(\bar d u)_{V,A}|0\rangle$ is to include more data in the numerical analysis.
Note that our derivation depends on the $SU(2)$ helicity $[SU(2)_h]$ symmetry at large $t$ ($t\to\infty$),
where quarks can be seen as massless particles. In the baryonic $B$ decay processes, 
since $t$ is ranging from $(m_{\bf B}+m_{\bf \bar B'})^2$ to $(m_B-m_M)^2$,
instead of $t\to\infty$, the fact that the quarks are no longer massless
can induce the $SU(2)_h$ symmetry breaking. Hence,
$\delta C_{||(\overline{||})}$ and $\delta \bar C_{||}$ are added in Eq.~(\ref{timelikeF3}) 
to estimate the possible broken symmetry effect.
One has derived $F_2=F_1/(t\text{ln}[t/\Lambda_0^2])$ in the pQCD model~\cite{Belitsky:2002kj},
which verifies the parameterization in Eq.~(\ref{timelikeF2}). By contrast,
the model calculation for $h_A$ has not been available yet.
In the $SU(3)$ flavor $[SU(3)_f]$ symmetry,
$C_{h_A}$ can be related as~\cite{Hsiao:2014zza}
\begin{eqnarray}\label{ChA} 
C_{h_A}=-\frac{1}{\sqrt 6}(C_D+3C_F)\,,\;\;\;C_{h_A}=C_D+C_F\,,
\end{eqnarray}
for $\langle \Lambda\bar p|(\bar s u)_A|0\rangle$ 
and $\langle n\bar p|(\bar d u)_A|0\rangle$, respectively.

To integrate over the phase space in the three-body $B\to{\bf B\bar B'}M$ decays,
we adopt the equation as~\cite{Geng:2006wz,Hsiao:2016amt,Chen:2003nz}
\begin{eqnarray}\label{Gamma}
\Gamma=\int^{+1}_{-1}
\int^{(m_B-m_M)^2}_{(m_{\bf B}+m_{\bf \bar B'})^2}
\frac{\beta_t^{1/2}\lambda^{1/2}_t}{(8\pi m_B)^3}
|\bar {\cal M}|^2\;dt\;d\text{cos}\theta\;,
\end{eqnarray}
where  $\beta_t=[1-(m_{\bf B}+m_{\bf \bar B'})^2/t][1-(m_{\bf B}-m_{\bf\bar B'})^2/t]$,
$\lambda_t=[(m_B+m_M)^2-t][(m_B-m_M)^2-t]$, and $\Gamma$ represents the decay width. 
Moreover, $|\bar {\cal M}|^2$ denotes the squared amplitude summed over the baryon spins.
We choose $\theta$ as the angle between $\bf \bar B'$ and $M$ moving directions
in the $\bf B\bar B'$ rest frame. Accordingly, 
the (anti-)baryon energy can be a function of $\cos\theta$,
given by
\begin{eqnarray}\label{Ep}
E_{\bf B}&=&\frac{t+m_B^2-m_\pi^2+\beta_t^{1/2}\lambda_t^{1/2}\cos\theta}{4m_B}\,,\nonumber\\
E_{\bf\bar B'}&=&\frac{t+m_B^2-m_\pi^2-\beta_t^{1/2}\lambda_t^{1/2}\cos\theta}{4m_B}\,.
\end{eqnarray}

We reduce $|\bar {M}|^2(B\to\Lambda\bar p \pi)$ as  
\begin{eqnarray}\label{amp2}
&&
|\bar {M}|^2(B\to\Lambda\bar p \pi)\simeq a+b\cos\theta+c\cos^2\theta\,,\nonumber\\
&&
a=2|\alpha_1|^2 F_{\pi1}^2\{F_1^2(m_B^2-t)^2+g_A^2[4m_p^2(2m_B^2-t)+(m_B^2-t)^2]\}\nonumber\\
&&+2|\alpha_6|^2(m_B^2/m_b)^2 F_{\pi0}^2[f_S^2(t-4m_p^2)+g_P^2 t]
-8\Re(\alpha_1\alpha_6^*)F_{\pi0}F_{\pi1}g_A g_Pm_p(m_B^4/m_b)\,,\nonumber\\
&&
b=-8\Re(\alpha_1\alpha_6^*)F_{\pi0}F_{\pi1}F_1f_S m_p(m_B^2/m_b)
[(t-4m_p^2)(m_B^2-t)^2/t]^{1/2}\,,\nonumber\\
&&
c=2|\alpha_1|^2 [(t-4m_p^2)(m_B^2-t)^2/t]F_{\pi1}^2(F_1^2+g_A^2)\,,
\end{eqnarray}
with $m_\Lambda-m_p\simeq 0$, $m_\pi/m_B\simeq 0$ and
$F_{\pi 0}-F_{\pi 1}\simeq 0$. Note that $|\bar {M}|^2(B\to\Lambda\bar p \pi)$ 
in the reduced form is for a simple presentation; however, 
no approximation is made in the real calculation.
Similarly, we obtain
\begin{eqnarray}
&&
|\bar {M}|^2(B\to\Lambda\bar p \rho)\simeq a^*+b^*\cos\theta+c^*\cos^2\theta\,,\nonumber\\
&&
a^*=|\alpha_1|^2(m_B^2-t)^2/(2m_\rho^2 m_B^2 t)
[A_1 m_B^2-A_2(m_B^2-t)]^2[F_1^2 t+g_A^2(t-4m_p^2)]\nonumber\\
&&
+2|\alpha_6|^2/m_b^2[A_0(m_B^2-t)]^2[f_S^2(t-4m_p^2)+g_P^2 t]\,,\nonumber\\
&&
b^*=4\Re(\alpha_1\alpha_6^*)m_p/(m_\rho m_b m_B)
[(t-4m_p^2)(m_B^2-t)^2/t]^{1/2} (m_B^2-t)\nonumber\\
&&
\times A_0[A_1 m_B^2-A_2(m_B^2-t)]F_1 f_S\,,\nonumber\\
&&
c^*=|\alpha_1|^2 /(2m_\rho^2 m_B^2)[(t-4m_p^2)(m_B^2-t)^2/t]
[A_1 m_B^2-A_2(m_B^2-t)]^2 (F_1^2+g_A^2)\,.
\end{eqnarray}
For the angular asymmetry,
we define
\begin{eqnarray}\label{AFB}
{\cal A}_{FB}\equiv\frac{\int^{+1}_0\frac{d\Gamma}{d\cos\theta}d\cos\theta
-\int^0_{-1}\frac{d\Gamma}{d\cos\theta}
d\cos\theta}{\int^{+1}_0\frac{d\Gamma}{d\cos\theta}
d\cos\theta+\int^0_{-1}\frac{d\Gamma}{d\cos\theta}
d\cos\theta}\,,
\end{eqnarray}
where $d\Gamma/d\cos\theta$ is the angular distribution.

\section{Numerical Analysis}
\begin{figure}[t!]
\centering
\includegraphics[width=2.8in]{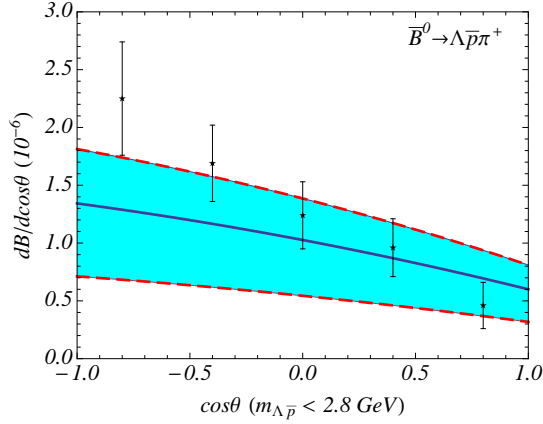}
\caption{The angular distribution of $\bar B^0\to \Lambda\bar p\pi^+$
with the solid (dotted) line for the central value (error), where 
the data points are adopted from Ref.~\cite{Belle:2007lbz}.}\label{fig2}
\end{figure}

In the numerical analysis, we take the CKM matrix elements as~\cite{pdg}
\begin{eqnarray}
(V_{ud},V_{us},V_{ub})
&=&(1-\lambda^2/2,\lambda,A\lambda^3(\rho-i\eta))\,,\nonumber\\
(V_{td},V_{ts},V_{tb})
&=&(A\lambda^3(1-\rho-i\eta),-A\lambda^2,1)\,,
\end{eqnarray}
with 
$(\lambda,A,\rho,\eta)=(0.2265,0.790,0.145\pm 0.017,0.366\pm 0.011)$
in the Wolfenstein parameterization. 
We use the mesonic form factors from Ref.~\cite{MS},
which can be found in Table~\ref{MF}. From Ref.~\cite{ali}, 
$c_i^{eff}\;(i=1,2, ..., 6)$ can lead to
\begin{eqnarray} 
\alpha_1&=&(-13.7-10.7i,-15.2-11.4i,-18.3 -12.7i)\,,\nonumber\\
\alpha_6&=&(47.5+6.4i,49.6+6.9i,53.7+7.7i)\,,
\end{eqnarray}
with $N_c=(2,3,\infty)$, where $N_c$ is taken as a floating number, in order that 
the non-factorizable QCD corrections can be estimated
in the generalized edition of the factorization~\cite{ali}.
%
\begin{table}[t!]
\caption{The form factors of $B\to \pi,\rho$ are adopted from Ref.~\cite{MS},
with $M_{A(B)}=5.32$~GeV and $(M_A,M_B)=(5.27,5.32)$~GeV 
for $\pi$ and $\rho$, respectively. In the second row, $F_{A}(0)$ and $F_{B}(0)$ correspond to 
$[F_{\pi1}(t),V_1(t),A_0(t)]$ and $[F_{\pi0}(t),A_{1,2}(t)]$, respectively,
with the zero value of the momentum transfer squared ($t=0$).}\label{MF}
{
\begin{tabular}{|c|cc|cccc|}
\hline
$B\to \pi,\rho$ &$F_{\pi0}$&$F_{\pi1}$&$V_1$&$A_0$&$A_1$&$A_2$\\\hline
$F_{A,B}(0)$   &0.29   &0.29   &0.31 &0.30 &0.26 &0.24\\
$\sigma_1$     &0.76   &0.48   &0.59 &0.54 &0.73 &1.40\\
$\sigma_2$     &0.28   &-----    &-----  &-----  &0.10 &0.50\\\hline
\end{tabular}}
\end{table}
%
In Eqs.~(\ref{timelikeF3}) and (\ref{ChA}),
there are totally eight constants that correspond to the baryonic form factors:
\begin{eqnarray}
C_{||}\,,\;
\delta C_{||}\,,\;
C_{\overline{||}}\,,\;
\delta C_{\overline{||}}\,,\;
\bar C_{||}\,,\;\delta \bar C_{||}\,,
C_D\,,\;C_F\,.
\end{eqnarray}
We perform the minimum $\chi^2$-fit to extract the constants, 
which includes the experimental inputs from 
${\cal B}(\bar B^0\to\Lambda\bar p\pi^+)$, ${\cal B}(B^-\to\Lambda\bar p\pi^0(\rho^0))$,
${\cal A}_{FB}(\bar B^0\to\Lambda\bar p\pi^+)$, and ${\cal A}_{FB}(B^-\to\Lambda\bar p\pi^0)$
in Table~\ref{tab1}, and the angular distribution of $\bar B^0\to\Lambda\bar p\pi^+$ in Fig.~\ref{fig2};
the branching fractions of 
$\bar B^0\to n\bar p D^{*+}$~\cite{CLEO:2000vce}, 
$\bar B^0\to\Lambda\bar p D^{(*)+}$~\cite{Chang:2015fja}, 
$B^-\to \Lambda\bar p$~\cite{LHCb:2016nbc}, 
$D_s^-\to n\bar p$~\cite{CLEO:2008aum,BESIII:2018cfe}, and 
${\cal A}_{FB}(\bar B^0\to \Lambda\bar p D^{(*)+})$~\cite{Chang:2015fja}
are also included.
We thus present the results of the global fit in Table~\ref{fit_result}.
%
\begin{table}[t!]
\caption{Fit results of the constants ($C_i$) derived from the baryonic form factors,
along with the $\chi^2$ value; 
$n.d.f$ denotes the number of degrees of freedom.}\label{fit_result}
\scriptsize
\begin{tabular}{|cr|}
\hline
$(\chi^2,n.d.f,C_i)$&Fit values$\;\;\;\;\;\;\;\;\;\;\;\;\;$
\\
\hline\hline
$\chi^2$&24.4$\;\;\;\;\;\;\;\;\;\;\;\;\;\;\;\;$
\\
$n.d.f$&9$\;\;\;\;\;\;\;\;\;\;\;\;\;\;\;\;\;\;$
\\
$C_{||}$&$(150.8\pm 5.7)\;{\rm GeV}^{4}$
\\
$\delta C_{||}$&$(31.9\pm 7.1)\;{\rm GeV}^{4}$
\\
$C_{\overline{||}}$&$(27.4\pm 27.3)\;{\rm GeV}^{4}$
\\
$\delta C_{\overline{||}}$&$(-735.0\pm 293.0)\;{\rm GeV}^{4}$
\\
$\bar C_{||}$&$(511.2\pm 74.4)\;{\rm GeV}^{4}$
\\
$\delta \bar C_{||}$&$(-317.8\pm 169.1)\;{\rm GeV}^{4}$
\\
$C_D$&$(-761.1\pm 128.0)\;{\rm GeV}^{6}$
\\
$C_F$&$(905.7\pm 119.8)\;{\rm GeV}^{6}$
\\
\hline
\end{tabular}
\end{table}
With the extracted constants, we calculate the branching fractions and angular asymmetries 
of $B\to\Lambda\bar p M$, and draw the angular distribution of $\bar B^0\to \Lambda\bar p\pi^+$ 
with $m_{\Lambda\bar p}<2.8$~GeV, which are given in Table~\ref{tab1} and  Fig.~\ref{fig2}, respectively.
%
\begin{table}[b!]
\caption{Branching fractions and angular asymmetries of 
the baryonic decay channels, where the first error of our results 
estimates the non-factorizable effects, 
while the second one combines the uncertainties from 
CKM matrix elements and the hadronic parameters.}\label{tab1}
{
\footnotesize
\begin{tabular}{|l|c|c|}
\hline 
$\;\;\;\;$ decay modes
&our results
&experimental data
\\\hline\hline
$10^{6}{\cal B}(\bar B^0\to \Lambda\bar p \pi^+)$    
&$3.2^{+0.6+2.5}_{-0.3-1.1}$
&$3.1\pm 0.3$~\text{\cite{Belle:2007lbz,pdg}}\\
$10^{6}{\cal B}(B^-\to \Lambda\bar p \pi^0)$
&$1.8^{+0.3+1.4}_{-0.2-0.6}$
&$3.0\pm 0.7$~\text{\cite{Belle:2007lbz,pdg}}\\
$10^{6}{\cal B}(\bar B^0\to \Lambda\bar p \rho^+)$
&$9.2^{+0.9+5.7}_{-1.9-3.5}$
&\\
$10^{6}{\cal B}(B^-\to \Lambda\bar p \rho^0)$
&$5.0^{+0.5+3.1}_{-1.0-1.9}$
&$4.8\pm 0.9$~\text{\cite{Belle:2009ixu,pdg}}\\
\hline
${\cal A}_{FB}(\bar B^0\to \Lambda\bar p \pi^+)$             
&$(-14.6^{+0.9}_{-1.5}\pm 6.9)\%$
&$(-41\pm 11\pm 3)\%$~\text{\cite{Belle:2007lbz}}\\
${\cal A}_{FB}(B^-\to \Lambda\bar p \pi^0)$           
&$(-14.6^{+0.9}_{-1.5}\pm 6.9)\%$
&$(-16\pm 18\pm 3)\%$~\text{\cite{Belle:2007lbz}}\\
${\cal A}_{FB}(\bar B^0\to \Lambda\bar p \rho^+)$             
&$(4.1^{+2.8}_{-0.7}\pm 2.0)\%$
&\\
${\cal A}_{FB}(B^-\to \Lambda\bar p \rho^0)$           
&$(4.1^{+2.8}_{-0.7}\pm 2.0)\%$
&\\
\hline
\end{tabular}}
\end{table}
%
%
%
\section{Discussions and Conclusions}
We study the penguin-dominant $B\to\Lambda\bar p \pi$ decay
with the branching fraction and angular asymmetry. 
With $\chi^2/n.d.f\simeq 2.7$ calculated from Table~\ref{fit_result}, it demonstrates that
the theoretical study can accommodate the experimental data.
Particularly, we find that $\Delta\chi^2=14.3$ 
that comes from ${\cal A}_{FB}(B\to\Lambda\bar p\pi)$ and five data points of 
$d{\cal B}(\bar B^0\to\Lambda\bar p\pi^+)/d\cos\theta$
gives sizeable contribution to the total $\chi^2$ value, indicating that 
more accurate measurements of the angular distribution (asymmetry)
can improve the fit.

In Eq.~(\ref{amp2}), 
$|\bar {M}|^2(B\to\Lambda\bar p \pi)\simeq a+b\cos\theta+c\cos^2\theta$
can be useful for our investigation.
It is found that $2|\alpha_6|^2(m_B^2/m_b)^2 F_{\pi0}^2[f_S^2(t-4m_p^2)+g_P^2 t]$
in the $a$ term gives the main contribution to the total branching fraction.
In the $b$ term, 
$-8\Re(\alpha_1\alpha_6^*)F_{\pi0}F_{\pi1}F_1f_S m_p(m_B^2/m_b)[(t-4m_p^2)(m_B^2-t)^2/t]^{1/2}$
is responsible for the angular asymmetry. However,
the $c$ term with $|\alpha_1|^2$ is insignificant.
As a consequence, we obtain 
${\cal B}(\bar B^0\to \Lambda\bar p \pi^+)=(3.2^{+0.6+2.5}_{-0.3-1.1})\times 10^{-6}$.
Besides, we obtain
${\cal A}_{FB}(\bar B^0\to \Lambda\bar p \pi^+)=(-14.6^{+0.9}_{-1.5}\pm 6.9)\%$
that has 2 standard deviation departure from the experimental value of $(-41\pm 11\pm 3)\%$.

Since we get $C_{g_P}=(0.38\pm 0.37)C_{f_S}$ different from 
$C_{g_P}=C_{f_S}$ in the $SU(2)$ helicity symmetry at $t\to\infty$,
it clearly indicates a broken symmetry effect with $\delta\bar C_{||}$.
Currently, $\delta\bar C_{||}$ is determined with a large uncertainty,
which reflects the fact that
${\cal A}_{FB}(B\to\Lambda\bar p\pi)$ and the angular distribution in Fig.~\ref{fig2}
have not been precisely measured. Without a model calculation, 
we obtain $C_{h_A}$ from the fit. It is found that
$C_{h_A}=-798.6$~GeV$^6$ for $\langle\Lambda\bar p|(\bar su)_A|0\rangle$
gives 2.8\% of ${\cal B}(\bar B^0\to \Lambda\bar p \pi^+)$ and
3.4\% of ${\cal A}_{FB}(\bar B^0\to \Lambda\bar p \pi^+)$.

The angular asymmetry of $\bar B^0\to\Lambda\bar p \pi^+$ decay 
was once studied in Ref.~\cite{Tsai},
where ${\cal A}_{FB}(\bar B^0\to\Lambda\bar p \pi^+)\simeq 0$ is not verified by the observation. 
The cause is that $g_P=f_S$ as a strong relation has been used,
such that $g_P$ with $2|\alpha_6|^2[f_S^2(t-4m_p^2)+g_P^2 t]$ in the $a$ term
becomes the dominant form factor in the branching fraction. By contrast,
$f_S$ turns out to be a less important form factor both in the $a$ and $b$ terms,
leading to ${\cal A}_{FB}(\bar B^0\to\Lambda\bar p \pi^+)\simeq 0$. 

For the first time, we study the angular asymmetry
of the charmless $B\to{\bf B\bar B'}M$ decay with $M$ as a vector meson. 
We predict ${\cal A}_{FB}(B^-\to \Lambda\bar p \rho^0)=           
(4.1^{+2.8}_{-0.7}\pm 2.0)\%$. It is interesting to note that
${\cal A}_{FB}(B\to \Lambda\bar p \rho)$ is not as large as 
${\cal A}_{FB}(B\to \Lambda\bar p \pi)$. This is due to $A_1\simeq A_2$,
which suppresses $4\Re(\alpha_1\alpha_6^*)m_p/(m_\rho m_b m_B)[(t-4m_p^2)(m_B^2-t)^2/t]^{1/2} (m_B^2-t)
A_0[A_1 m_B^2-A_2(m_B^2-t)]F_1 f_S$ in the $b^*$ term,
resulting in a suppressed angular asymmetry.

We find no source to violate the isospin relation.
Since $\pi^0(\rho^0)=(u\bar u-d\bar d)/\sqrt 2$ and $\pi^+(\rho^+)=u\bar d$,
it results in $\sqrt 2\langle \pi^0(\rho^0)|(\bar u b)|B^-\rangle
=\langle \pi^+(\rho^+)|(\bar u b)|\bar B^0\rangle$~\cite{Hsiao:2019ann,Chua:2002yd}.
We hence obtain 
\begin{eqnarray}
{\cal B}(\bar B^0\to \Lambda\bar p \pi^+(\rho^+))&\simeq& 2 {\cal B}(B^-\to \Lambda\bar p \pi^0(\rho^0))\,,\nonumber\\
{\cal A}_{FB}(\bar B^0\to \Lambda\bar p \pi^+(\rho^+))&=&{\cal A}_{FB}(B^-\to \Lambda\bar p \pi^0(\rho^0))\,,
\end{eqnarray}
which suggests 
${\cal B}(\bar B^0\to \Lambda\bar p \rho^+)\simeq 10^{-5}$ that
has not been measured yet.

In summary, we have investigated the angular asymmetry of $B\to\Lambda\bar p M$.
In particular, we have obtained
${\cal A}_{FB}(\bar B^0\to \Lambda\bar p \pi^+)=(-14.6^{+0.9}_{-1.5}\pm 6.9)\%$
with 2 standard deviation departure from the experimental value of $(-41\pm 11\pm 3)\%$.
We have hence reduced the deviation caused by 
${\cal A}_{FB}(\bar B^0\to \Lambda\bar p \pi^+)\simeq 0$
previously studied in the literature. We have calculated
${\cal A}_{FB}(B^-\to \Lambda\bar p \rho^0)=         
(4.1^{+2.8}_{-0.7}\pm 2.0)\%$, which can be the first prediction
for the charmless $B\to{\bf B\bar B'}M$ decay with $M$
as a vector meson. According to the isospin relation,
it has been calculated that ${\cal B}(\bar B^0\to \Lambda\bar p \rho^+)
=(9.2^{+0.9+5.7}_{-1.9-3.5})\times 10^{-6}$, promising to be measured
by the LHCb and Belle~II experiments.

\section*{ACKNOWLEDGMENTS}
YKH would like to thank Prof.~Yao Yu for useful discussions.
YKH was supported in part by NSFC (Grants No.~11675030 and No.~12175128).
LS was supported in part by NSFC (Grant No.~12061141006)
and Joint Large-Scale Scientific Facility Funds of the NSFC and CAS (Grant No.~U1932108).


\end{document}